\documentclass[aps,prl,twocolumn,superscriptaddress,showpacs,
  preprintnumbers,amsmath,amssymb]{revtex4}
\usepackage{graphicx}
\usepackage{dcolumn}
\graphicspath{{ps}}

\begin{document}

\title{ \quad\\[1.0cm] \boldmath Measurement of Branching Fraction and
  Direct $CP$ Asymmetry in $B^0 \to \rho^0 \pi^0$ Decays}

\affiliation{Budker Institute of Nuclear Physics, Novosibirsk}
\affiliation{Chiba University, Chiba}
\affiliation{Chonnam National University, Kwangju}
\affiliation{University of Cincinnati, Cincinnati, Ohio 45221}
\affiliation{University of Hawaii, Honolulu, Hawaii 96822}
\affiliation{High Energy Accelerator Research Organization (KEK), Tsukuba}
\affiliation{Institute of High Energy Physics, Chinese Academy of Sciences, Beijing}
\affiliation{Institute of High Energy Physics, Vienna}
\affiliation{Institute of High Energy Physics, Protvino}
\affiliation{Institute for Theoretical and Experimental Physics, Moscow}
\affiliation{J. Stefan Institute, Ljubljana}
\affiliation{Kanagawa University, Yokohama}
\affiliation{Korea University, Seoul}
\affiliation{Kyungpook National University, Taegu}
\affiliation{Swiss Federal Institute of Technology of Lausanne, EPFL, Lausanne}
\affiliation{University of Ljubljana, Ljubljana}
\affiliation{University of Maribor, Maribor}
\affiliation{University of Melbourne, Victoria}
\affiliation{Nagoya University, Nagoya}
\affiliation{Nara Women's University, Nara}
\affiliation{National Central University, Chung-li}
\affiliation{National United University, Miao Li}
\affiliation{Department of Physics, National Taiwan University, Taipei}
\affiliation{H. Niewodniczanski Institute of Nuclear Physics, Krakow}
\affiliation{Niigata University, Niigata}
\affiliation{Nova Gorica Polytechnic, Nova Gorica}
\affiliation{Osaka City University, Osaka}
\affiliation{Osaka University, Osaka}
\affiliation{Panjab University, Chandigarh}
\affiliation{Peking University, Beijing}
\affiliation{Princeton University, Princeton, New Jersey 08544}
\affiliation{RIKEN BNL Research Center, Upton, New York 11973}
\affiliation{Saga University, Saga}
\affiliation{University of Science and Technology of China, Hefei}
\affiliation{Seoul National University, Seoul}
\affiliation{Shinshu University, Nagano}
\affiliation{Sungkyunkwan University, Suwon}
\affiliation{University of Sydney, Sydney NSW}
\affiliation{Tata Institute of Fundamental Research, Bombay}
\affiliation{Toho University, Funabashi}
\affiliation{Tohoku Gakuin University, Tagajo}
\affiliation{Tohoku University, Sendai}
\affiliation{Department of Physics, University of Tokyo, Tokyo}
\affiliation{Tokyo Institute of Technology, Tokyo}
\affiliation{Tokyo Metropolitan University, Tokyo}
\affiliation{Tokyo University of Agriculture and Technology, Tokyo}
\affiliation{University of Tsukuba, Tsukuba}
\affiliation{Virginia Polytechnic Institute and State University, Blacksburg, Virginia 24061}
\affiliation{Yonsei University, Seoul}
   \author{J.~Dragic}\affiliation{High Energy Accelerator Research Organization (KEK), Tsukuba} 
   \author{K.~Abe}\affiliation{High Energy Accelerator Research Organization (KEK), Tsukuba} 
   \author{K.~Abe}\affiliation{Tohoku Gakuin University, Tagajo} 
   \author{I.~Adachi}\affiliation{High Energy Accelerator Research Organization (KEK), Tsukuba} 
   \author{H.~Aihara}\affiliation{Department of Physics, University of Tokyo, Tokyo} 
   \author{D.~Anipko}\affiliation{Budker Institute of Nuclear Physics, Novosibirsk} 
   \author{Y.~Asano}\affiliation{University of Tsukuba, Tsukuba} 
   \author{T.~Aushev}\affiliation{Institute for Theoretical and Experimental Physics, Moscow} 
   \author{S.~Bahinipati}\affiliation{University of Cincinnati, Cincinnati, Ohio 45221} 
   \author{A.~M.~Bakich}\affiliation{University of Sydney, Sydney NSW} 
% ...
\collaboration{The Belle Collaboration}

\begin{abstract}
  We report a measurement of the branching fraction of the decay $B^0 \to \rho^0 \pi^0$, using
  $386 \times 10^6$ $B\overline{B}$ pairs collected at the $\Upsilon(4S)$
  resonance with the Belle detector at the KEKB asymmetric-energy $e^+e^-$    
  collider. We detect $51^{+14}_{-13}$ signal events with a
  significance
  of 4.2 standard deviations, including systematic
  uncertainties, and measure the branching fraction to be
  ${\mathcal B}\left( B^0 \to \rho^0\pi^0 \right) =                           
  \left( 3.12^{+0.88}_{-0.82}({\rm stat}) \pm 0.33 ({\rm syst}) ^{+0.50}_{-0.68} ({\rm model}) \right)        
  \times 10^{-6}.$
  We also perform the first measurement of direct
  $CP$ violating asymmetry in this mode.
\end{abstract}


\maketitle

\tighten

    {\renewcommand{\thefootnote}{\fnsymbol{footnote}}}
    \setcounter{footnote}{0}

Tests of the Kobayashi-Maskawa model~\cite{ref:km} for $CP$ violation are ongoing.
In particular, $B$-factories are directing focus towards measurements of the
lesser known angles of the Cabibbo-Kobayashi-Maskawa (CKM) triangle,
$\phi_2$ and $\phi_3$.
Measurements of $\phi_2$ typically rely on time-dependent $CP$-violation studies
of $B$ meson decays to $\pi^+ \pi^-$, $\rho^\pm \pi^\mp$ and $\rho^+ \rho^-$~\cite{ref:time-dep,ref:chinchi},
since the leading tree amplitudes for these processes involve the relevant 
CKM phases. However, penguin amplitudes may also contribute
significantly in these decays and --- via introducing additional
unknown phases ---
greatly impair $\phi_2$ constraints from the time-dependent
measurements. In such cases, isospin analyses can be  
employed to separate the tree-level process from penguin
contamination~\cite{ref:isospin}.

Measurements of $\phi_2$ from the $\rho\pi$ system
rely on knowledge of the $B^0 \to \rho^0\pi^0$ branching
fraction~\cite{ref:isospin,ref:rhopi_dalitz}.                        
Since the tree amplitude of $B^0 \to \rho^0\pi^0$ decay is color
suppressed, the decay rate is sensitive to the penguin amplitude
contribution.
Thus, the $\rho^0\pi^0$ branching fraction plays a critical role
in constraining the $\phi_2$ uncertainty due to penguin pollution 
from time-dependent $B^0 \to \rho^\pm\pi^\mp$
measurements~\cite{ref:chinchi,ref:isospin}.
Furthermore, measurement of $\phi_2$ from the full $B \to \rho\pi$
isospin analysis requires the $\rho^0\pi^0$ branching fraction along
with its $CP$ asymmetry. Since the  branching fractions and $CP$
asymmetries of all the other $\rho\pi$ final states have been
measured~\cite{ref:rhopi_bfs}, $\rho^0\pi^0$ is the only channel
that remains to complete the isospin pentagons. A simplification,
whereby the pentagons collapse into quadrangles, is also possible
if the $\rho^0 \pi^0$ amplitude is sufficiently small.

An alternative technique to measure $\phi_2$ from the
$\rho\pi$ system, even if penguin contamination is large, is a
time-dependent amplitude analysis of $B^0 \to \pi^+\pi^-\pi^0$~\cite{ref:rhopi_dalitz}.                           
Here, the interferences between $\rho^+ \pi^-$, $\rho^0 \pi^0$ and
$\rho^- \pi^+$ contributions to the $\pi^+\pi^-\pi^0$ final state
provide the critical information on the unknown phases introduced by penguin amplitudes. 
Recently, the first time-dependent studies of the $\pi^+\pi^-\pi^0$
Dalitz plot have been performed~\cite{ref:babar_rhopi_dalitz}. In 
these studies, a simplification is made with the assumption that the 
$\rho^0\pi^0$ contribution is small. A more complex
time-dependent Dalitz analysis is required if this is not the case.

The Belle Collaboration reported first evidence of the $B^0 \to \rho^0
\pi^0$ decay~\cite{ref:rho0pi0_evidence} with a branching fraction
larger than most predictions~\cite{ref:rho0pi0_predictions}, and a
central value above the 90\% confidence-level 
upper limit set by the BaBar Collaboration~\cite{ref:rho0pi0_ul}.
In this paper, we report an improved measurement of the 
$B^0 \to \rho^0 \pi^0$ branching fraction~\cite{ref:cc}, using $2.5$ times more
data, and perform a first direct $CP$ violation search in this mode.
The results are consistent with and supersede those reported in our
previous publication.
The analysis is based on $(385.8 \pm 4.8) \times 10^6$
 $B\overline{B}$ pairs, 
collected  with the Belle detector at the KEKB asymmetric-energy
$e^+e^-$ collider~\cite{ref:KEKB} that operates at the $\Upsilon(4S)$ resonance.
The production rates of $B^+B^-$ and $B^0\overline{B}{}^0$ pairs
are assumed to be equal.

The Belle detector~\cite{ref:Belle,ref:svd2}
is a large-solid-angle magnetic spectrometer that 
consists of a silicon vertex detector,
a 50-layer central drift chamber (CDC), an array of
aerogel threshold Cherenkov counters (ACC), 
a barrel-like arrangement of time-of-flight
scintillation counters (TOF), and an electromagnetic calorimeter
comprised of CsI(Tl) crystals (ECL) located inside 
a superconducting solenoid coil that provides a 1.5~T
magnetic field.  An iron flux-return located outside of
the coil is instrumented to detect $K_L^0$ mesons and to identify
muons.

The $B$ reconstruction procedure is identical to our
previously published analysis~\cite{ref:rho0pi0_evidence}.
Charged tracks are required to originate from the interaction point
and have transverse momenta greater than $100~{\rm MeV}/c$.
Pions are identified by combining information
from the ACC, TOF and the CDC $dE/dx$ measurements.
We further reject tracks that are consistent with an electron
hypothesis.
Pairs of photons with invariant masses in the range
$0.115\ {\rm GeV}/c^2 < m_{\gamma\gamma} < 0.154\ {\rm GeV}/c^2$ are
used to form the $\pi^0$ mesons. 
The photon energy in the laboratory frame is required to be greater than
$50~(100)~{\rm MeV}$ in the barrel (endcap) region of the ECL.
The $\pi^0$ candidates are required to have transverse momenta
greater than $100~{\rm MeV}/c$ in the laboratory frame and 
a loose requirement is made on $\chi^2_{\pi^0}$,
the goodness of fit of a $\pi^0$ mass-constrained fit of
the two photons. We also veto possible contributions to $\pi^+\pi^-\pi^0$ 
from charmed ($b \to c$) decays: $B^0 \to D^- \pi^+$, $\overline{D}{}^0
\pi^0$ and $J/\psi \pi^0$.

Signal $B$ candidates are identified with two kinematic variables:                                    
the beam-energy constrained mass                                                                     
$M_{\rm bc}\equiv \sqrt{E^2_{\rm beam}/c^4-p^2_B/c^2}$
and the energy difference $\Delta E \equiv E_B - E_{\rm beam}$.                               
Here, $E_B$ ($p_B$) is the reconstructed energy (momentum)
of the $B$ candidate, and $E_{\rm beam}$ is the beam energy,
all expressed in the centre-of-mass (CM) frame.                                        
We consider candidate events in the region                                                    
$-0.2\ {\rm GeV} < \Delta E < 0.4\ {\rm GeV}$ and                                           
$ M_{\rm bc} >  5.23\ {\rm GeV}/c^2$; and                                   
define signal regions in $\Delta E$ and $M_{\rm bc}$ as                                    
$-0.135\ {\rm GeV} < \Delta E < 0.082\ {\rm GeV}$ and                                         
$5.269\ {\rm GeV}/c^2 < M_{\rm bc} < 5.290\ {\rm GeV}/c^2$.                                                                                 
To select $\rho^0\pi^0$ from the 
$\pi^+\pi^-\pi^0$ candidates, we require                                                      
the $\pi^+\pi^-$ invariant mass to be in the range                                            
$0.5\ {\rm GeV}/c^2 < m_{\pi^+\pi^-} < 1.1\ {\rm GeV}/c^2$ and                              
the $\rho^0$ helicity angle to satisfy                                                        
$\left| \cos \theta^{\rho}_{\rm hel} \right| > 0.5$,                                          
where $\theta^{\rho}_{\rm hel}$ is                                                            
defined as the angle between the negative pion direction and the                              
opposite of the $B$ direction in the $\rho$ rest frame.                       
We explicitly veto contributions from $B^0 \to \rho^{\pm}\pi^{\mp}$ 
by the requirement                                                                   
$m_{\pi^{\pm}\pi^0} > 1.1\ {\rm GeV}/c^2$.                             
This requirement also vetoes the region of the Dalitz plot                                    
where the interference between $\rho^0\pi^0$ and $\rho^{\pm}\pi^{\mp}$ is strongest.                                
After all selection requirements, 11\% of
events have more than one candidate. Among those candidates the one 
with the smallest $\chi^2_{\rm vtx}/{ ndf} + \chi^2_{\pi^0}/{
  ndf}$ is selected,
where $\chi^2_{\rm vtx}$ is 
the goodness of fit of a vertex-constrained fit of $\pi^+\pi^-$.
                                                                                              
The dominant background originates from continuum
$e^+e^-\to q\overline{q}$ ($q = u, d, s, c$) production.
To separate the jet-like $q\overline{q}$ events,
we use event shape variables:
five modified Fox-Wolfram moments~\cite{ref:SFW}, combined
into a Fisher discriminant. We further combine the cosine
of the $B$ meson flight direction in the CM system with 
the output of the Fisher discriminant 
into a signal/background likelihood variable, ${\mathcal L}_{s/b}$,
and define the likelihood ratio                                         
${\mathcal R}={\mathcal L}_s/\left({\mathcal L}_s + {\mathcal L}_b \right)$.                  
Additional discrimination against continuum is achieved through                               
use of the $b$-flavour tagging algorithm~\cite{ref:tagging}.
We use the parameter $r$, with values between $0$ and $1$,
as a measure of the confidence that the remaining particles in the                    
event (other than $\pi^+\pi^-\pi^0$) originated from a flavour specific $B$                    
meson decay and --- as a corollary --- not from a continuum process.                           
                                                                                              
\begin{figure*}[htb]
  \hbox to \hsize{
    \includegraphics[width=0.201\hsize]{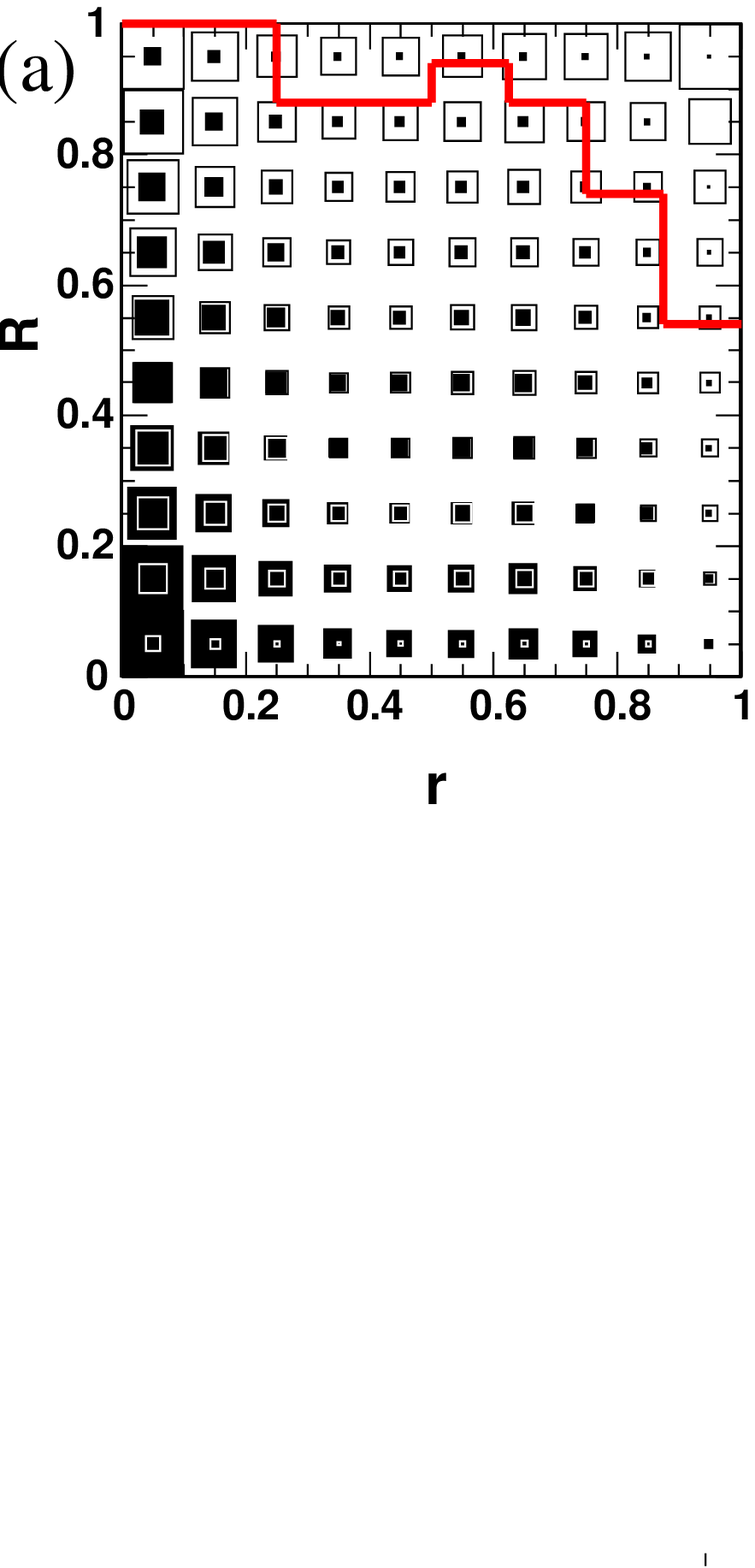}
    \includegraphics[width=0.399\hsize]{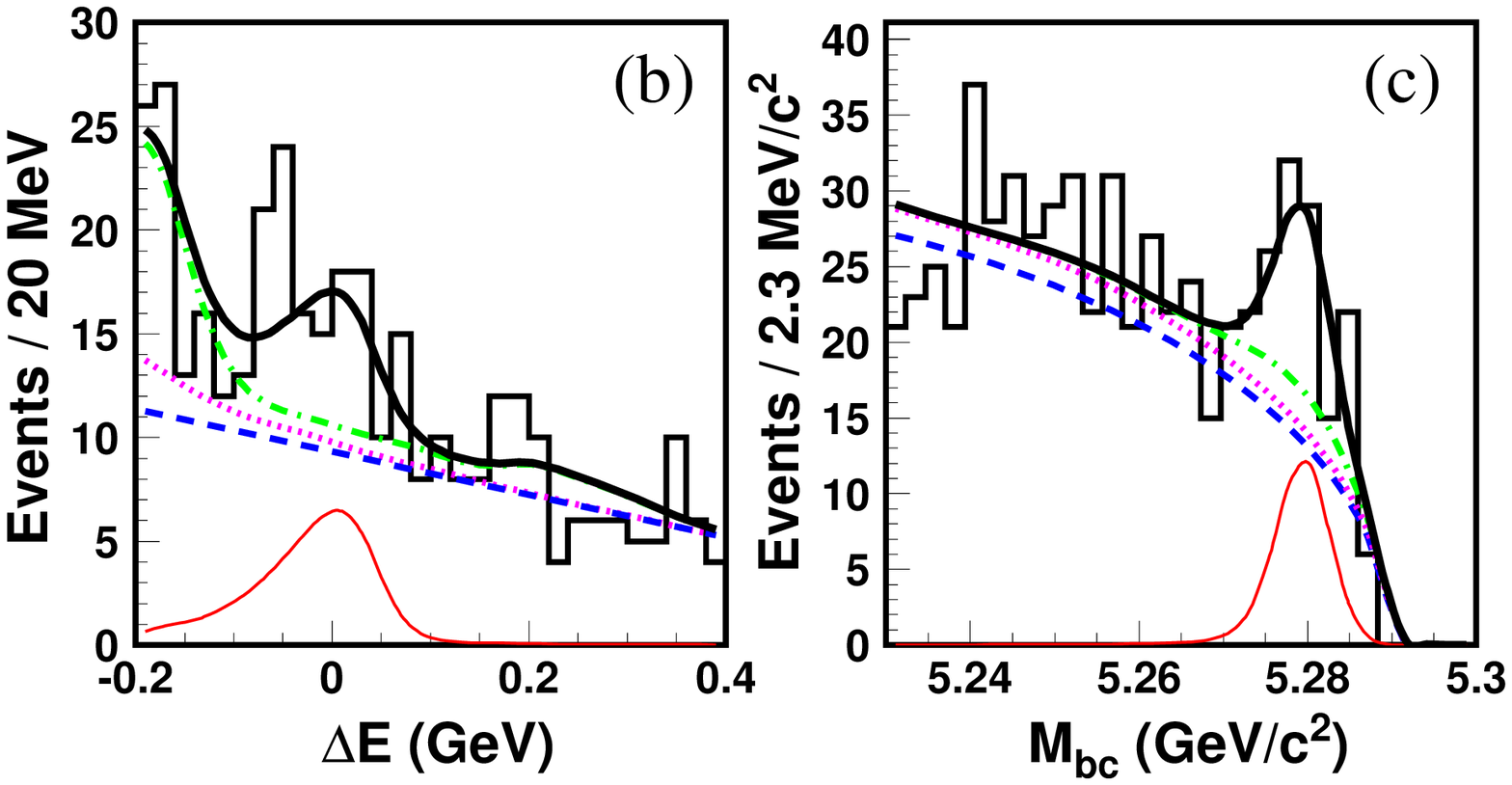}
    \includegraphics[width=0.399\hsize]{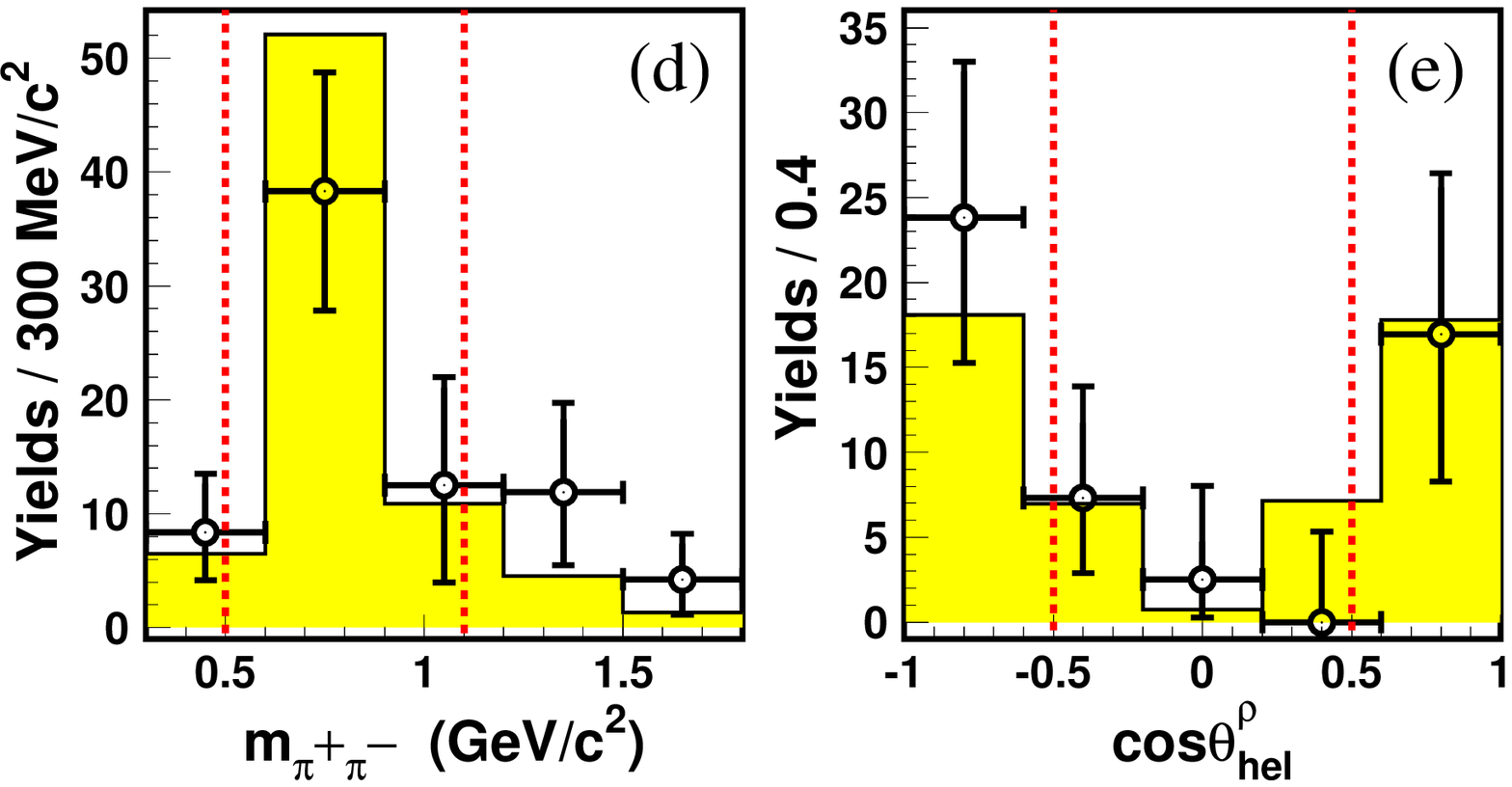}         
  }                                                                                           
  \caption{(a) Distribution of signal (continuum) 
    events in $r$-${\cal R}$ space shown with open (shaded) proportional boxes; the marked region (top-right)
    indicates the area selected. (b) (c) Distribution of $\Delta E$($M_{\rm bc}$) in the signal 
    region of $M_{\rm bc}$($\Delta E$).                                                       
    Projection of the fit result is shown as the thick solid curve;         
    the thin solid line represents the signal component;                                
    the dashed, dotted and dash-dotted curves represent, respectively, the
    cumulative background components from continuum processes,
    $b \to c$ decays, and charmless $B$ backgrounds. (d), (e)
    Distributions of fit yields                        
    in $m_{\pi^+\pi^-}$ and $\cos \theta^{\rho}_{\rm hel}$ variables                          
    for $\rho^0\pi^0$ candidate events.                                                       
    Points with error bars represent data fit results, and the histograms                     
    show signal MC expectation; the selection requirements described                          
    in the text are shown as dashed lines.                                                    
  }                                                                                           
  \label{fig:rho0pi0_main}                                                                     
\end{figure*}

We use an iterative procedure to find the optimal contiguous area in                          
$r$-${\cal R}$ space by maximising $N_s/\sqrt{N_s+N_b}$, where $N_s$
($N_b$) is the expected number of signal (background) events in
the $\Delta E$ and $M_{\rm bc}$ signal regions. Here, the
optimisation procedure 
assumes a branching fraction for
$B^0\to \rho^0\pi^0$ of $3.3 \times 10^{-6}$~\cite{ref:average}.
Anticipating the use of $r$ for its                   
primary purpose of flavour tagging in $CP$                        
asymmetry fits, the borders of the contiguous                  
area were constrained to match the six $r$ bins                  
employed in previous analyses.
The result of the optimisation procedure is that we select events within
the region shown in Fig.~\ref{fig:rho0pi0_main}(a).
We find 1397 candidates remain in the data.

We obtain the signal yield using an extended unbinned maximum-likelihood fit                           
to the $\Delta E$-$M_{\rm bc}$ distribution                                                   
of the selected candidate events.
The likelihood function is defined as
\begin{equation}
  {\cal L} = \exp \left( - \sum_{j,l} N_{j,l} \right)
  \prod_{i} \left( \sum_{j,l} N_{j,l} {\cal P}^i_{j} \right).
\end{equation}
Here, the index $i$ is the event identifier; 
$l$ distinguishes events in various $r$ bins; 
and $j$ runs over all six components included 
in the fitting function --- one for the signal, and the others for
continuum, $b \to c$ combinatorial, and the charmless $B$ backgrounds:
$B^+ \to \rho^+ \rho^0$, $B^+ \to \rho^+ \pi^0$ and $B^+ \to \pi^+
\pi^0$. $N_{j,l}$ represents the number of events, and
${\cal P}^i_{j} =  P_{j}(M^i_{{\rm bc}}, \Delta E^i)$ are 
two-dimensional probability density functions (PDFs).

The PDFs for signal, $b \to c$ and charmless $B$
backgrounds are taken from  smoothed two-dimensional histograms obtained
from Monte Carlo (MC) simulations.                           
For the $B^+ \to \rho^+ \rho^0$ channel, we assume a $100\%$               
longitudinally polarised decay~\cite{ref:rhoprho0_bf}.                                             
Small corrections to MC peak positions                                    
and widths
are applied to the signal PDF. These factors                               
are derived from control samples of reconstructed decays
$B^0 \to D^{*-}\rho^+$                                                                  
($D^{*-} \to \overline{D}{}^0\pi^-$, $\overline{D}{}^0 \to K^+\pi^-$; $\rho^+ \to \pi^+\pi^0$)
and                                                                                           
$B^+ \to \overline{D}{}^0\rho^+$                                                                 
($\overline{D}{}^0 \to K^+\pi^-$; $\rho^+ \to \pi^+\pi^0$),                                           
in which we require that the $\pi^0$ momentum be greater than                                 
$1.8\ {\rm GeV}/c$ in order to mimic the high momentum $\pi^0$ in our signal.                 
The two-dimensional PDF for the continuum background is described as                          
the product of a first-order polynomial in $\Delta E$                                         
and an ARGUS function~\cite{ref:argus} in $M_{\rm bc}$.                                          
All of the shape parameters describing the continuum background are free
parameters in the fit.
The normalisations of $B^+ \to \rho^+\pi^0$
($21.7 \pm 4.4$ events)
and                        
$B^+ \to \pi^+\pi^0$
($21.0 \pm 5.5$ events)
are fixed in the fit                              
according to previous measurements~\cite{ref:rhopi_bfs,ref:PDG},
and that of $b \to c$ background
($62 \pm 62$ events)
according to MC
expectation (assigning a conservative error); 
the normalisations of all other components are allowed to float.                        
                                                                                              
The fit result is shown in Fig.~\ref{fig:rho0pi0_main}(b) and (c).                             
The signal yield is found to be $50.9^{+14.3}_{-13.4}$ with $4.5 \sigma$ significance.                
The significance is defined as                                                                
$\sqrt{-2\ln({\mathcal L}_0/{\mathcal L}_{\rm max})}$,                                        
where ${\mathcal L}_{\rm max}$ (${\mathcal L}_0$)                                             
denotes the likelihood with the signal yield at its nominal value                             
(fixed to zero).                                                                              
The contribution from $B^+ \to \rho^+ \rho^0$ decays (which peaks in
the low $\Delta E$ region) is obtained from the 
fit as $43.1^{+13.2}_{-12.1}$ events; this value is  consistent with the MC
expectation ($33.9^{+8.1}_{-9.8}$ events) based on our branching fraction   
measurement of $B^+ \to \rho^+\rho^0$~\cite{ref:rhoprho0_bf}.         
A possible contribution from $B\to \omega(\pi^+\pi^-\pi^0) \pi^0$ decays
is also accounted for by floating the $B^+ \to \rho^+ \rho^0$ PDF, since the
two decays have similar distributions in $\Delta E$ and $M_{\rm bc}$.
To verify that the signal candidates originate from                                   
$B^0 \to \rho^0\pi^0$ decays,                                                                 
we change the criteria on $m_{\pi^+\pi^-}$ and $\cos \theta^{\rho}_{\rm hel}$                 
in turn, and repeat fits to the $\Delta E$-$M_{\rm bc}$ distribution.                         
The yields obtained in each $m_{\pi^+\pi^-}$ and $\cos \theta^{\rho}_{\rm hel}$               
bin are shown in Fig.~\ref{fig:rho0pi0_main}(d) and (e).                                       
                                                                                              
The $\cos \theta^{\rho}_{\rm hel}$ distribution is used to limit contributions                 
from $B^0 \to \sigma \pi^0$, $f_0(980)\pi^0$, $\eta^{'} \pi^0$, $K_S
\pi^0$ and $\pi^+\pi^-\pi^0$ (nonresonant),                     
which are expected to be flat in this variable.                                   
We perform a $\chi^2$ fit including components for                                            
pseudoscalar $\to$ pseudoscalar vector                                                        
($PV \sim \cos^2 \theta^{\rho}_{\rm hel}$), and                                               
pseudoscalar $\to$ pseudoscalar scalar ($PS \sim$ flat) decays,                               
for which the shapes are obtained from our $\rho^0\pi^0$ signal MC,                           
and a sample of $\sigma \pi^0$ MC~\cite{ref:sigma_pi0_mc}, respectively.                      
We also include a linear term to allow for possible interference.
We find that the $PS$ level is consistent with zero;
taking its uncertainty into account, we assign a model
error of $^{+0.0}_{-15.0}\%$ to the $PV$ component.                                                                 
The $m_{\pi^+\pi^-}$ distribution is consistent with the
expectation from $B^0 \to \rho^0\pi^0$ production.
                                                                                              
To extract the branching fraction,                                                            
we determine the reconstruction efficiency, $(4.99 \pm 0.03)\%$, from
MC and correct for small differences
between data and MC in the pion identification                                  
and continuum suppression requirements.                                                       
The correction factor due to charged pion identification ($0.872$) is obtained                         
in bins of track momentum and polar angle from an inclusive $D^*$ control                     
sample ($D^{*-} \to \overline{D}{}^0 \pi^-$, $\overline{D}{}^0 \to K^+ \pi^-$). The                         
corresponding systematic error is $\pm 3.1\%$.                                                    
For the continuum suppression requirement on $r$ and ${\mathcal R}$,                          
we use the control sample $B^0 \to D^{-}\rho^+$ 
($D^{-} \to K^+\pi^-\pi^-$; $\rho^+ \to \pi^+\pi^0$) to obtain                                           
an efficiency correction factor of $0.972$ and
a corresponding systematic error of $\pm 6.0\%$.                                                                 

We calculate additional systematic errors from the following sources:                            
PDF shapes by varying parameters by $\pm 1 \sigma$ ($^{+0.9}_{-2.0}\%$);                      
$\pi^0$ reconstruction efficiency by comparing the yields of                     
$\eta \to \pi^0 \pi^0 \pi^0$ and $\eta \to \gamma \gamma$ between data
and MC ($\pm 4.0\%$);               
track finding efficiency from a study of partially reconstructed $D^*$
decays ($\pm 2.4\%$);                                             
and data-MC efficiency differences due to
the $\Delta E > -0.2 \ {\rm GeV}$
requirement ($\pm 2.0\%$).                                                          
We repeat the fit after changing the                                                          
normalisation of the fixed background components according to the 
given errors
and obtain a systematic error of $^{+2.0}_{-1.8}\%$.
Using a large MC sample, the total systematic error from
possible charmless $B$ decays not otherwise included, $B^0 \to
K^{*0}\pi^0$ ($5.4\%$), $B^+ \to K^{*+}\pi^0$ ($1.5\%$) and $B^0 \to
K^+\rho^-$ ($0.5\%$), is $\pm 5.6\%$.                                                            

When the normalisations of all the                            
backgrounds fixed in the fit are simultaneously increased by $1\sigma$,                   
 the statistical significance decreases from $4.5 \sigma$ to $4.2 \sigma$;                    
we interpret the latter value as the significance of our result.                              
Finally, we estimate the uncertainty due to possible interference                  
with $B^0 \to \rho^\pm \pi^\mp$ by varying the $m_{\pi^\pm \pi^0}$ veto                       
requirement from $m_{\pi^\pm \pi^0} > 0 ~\rm MeV/c^2$ (no veto) to
$m_{\pi^\pm \pi^0} > 1.7 ~\rm GeV/c^2$.                                                  
We find the largest change in the result to be within $\pm 16\%$, and 
we include this value in the model error, so that the obtained 
$B^0 \to \rho^0\pi^0$ branching fraction is
$$                                                                                        
  {\mathcal B} =                                            
  \left( 3.12^{+0.88}_{-0.82}({\rm stat}) \pm 0.33 ({\rm syst}) ^{+0.50}_{-0.68} ({\rm model}) \right)        
  \times 10^{-6}.                                                                             
$$                                                                                            

Having observed a significant $B^0 \to \rho^0\pi^0$ signal, we utilize the
$B^0$/$\overline{B}{}^0$ separation provided by the flavour tagging to
measure the $CP$ asymmetry. For this purpose we replace
$\mathcal{P}^i_{j}$ of Eq. (1) with the expression
\begin{eqnarray} \mathcal{P}^i_{j,l} & = &\frac{1}{2}\left[ 1+ q^i
\cdot ({\cal A}^{'}_{CP})_{j,l} \right] P_{j}(M^i_{{\rm bc}}, \Delta E^i),
\end{eqnarray}
in which the indices keep the same meaning.
In this equation, $q$ represents the $b$-flavour charge
[$q = \: +1 \: (-1)$ when the tagging $B$ meson is a
  $B^0$ ($\overline{B}{}^0$)] and ${\cal A}^{'}_{CP}$ denotes
the effective charge asymmetry, such that $({\cal A}^{'}_{CP})_{j,l}=({\cal A}_{CP})_j (1-2\chi_d)(1-2 w_l)$.
Here, $({\cal A}_{CP})_j$ are the charge asymmetries
for the signal and the background components.
Further, $\chi_d=0.182\pm0.015$ \cite{ref:HFAG}
is the time-integrated mixing parameter and $w_l$ is the 
wrong-tag fraction. For continuum background, $\chi_d$ and $w_l$ are set to zero.
The data is divided into the six $r$-bins,
and the $r$-dependent wrong-tag fractions, $w_l$ ($l=1,\ldots,6$),
are determined using a high statistics sample of self-tagged
$B^0 \to D^{(*)-} \pi^+, D^{*-} \rho^+$ and $D^{*-} \ell^+ \nu$
events \cite{ref:tagging}.

\begin{figure}[htb]                                                                               
  \hbox to \hsize{                                                                            
    \includegraphics[width=0.98\hsize]{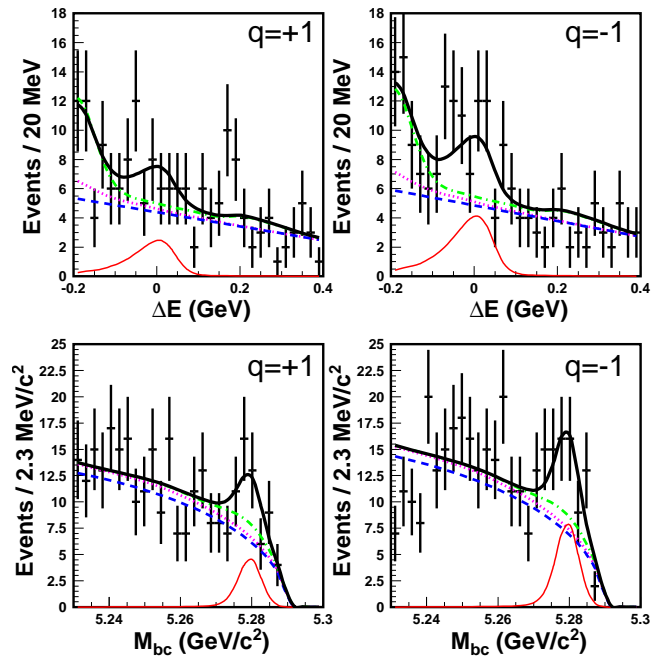}
  }
  \caption{$\Delta E$ and $M_{\rm bc}$ distributions (with projections of the fit
    results) shown separately for events tagged as $q=+1$ (left) and
    $q=-1$ (right).}                                                        
  \label{fig:rho0pi0_tagged}                                                                     
\end{figure}

The total number of signal, continuum background and $\rho^+\rho^0$ events are 
free parameters in the fit, and the remaining background components
(from $b \to c$, $\rho^+\pi^0$ and $\pi^+\pi^0$ decays) are fixed.
Also, the relative fractions for the signal and continuum background
components in different $r$ bins are allowed to float in the fit;
for the $b \to c$ and charmless $B$ decay backgrounds, they are fixed.
The only free ${\cal A}_{CP}$ parameter in the nominal fit 
is that of our signal; the others are fixed to be zero (for continuum and
$b \to c$) or at their previously measured values (for charmless $B$ backgrounds)~\cite{ref:PDG}.
We measure the direct $CP$ asymmetry in $B^0 \to \rho^0\pi^0$ decays to be
$$
 {\cal A}_{CP} = -0.53^{+0.67}_{-0.84}({\rm stat})^{+0.10}_{-0.15}({\rm syst}).
$$
The impact of background asymmetry ($^{+0.058}_{-0.127}$)
is the largest contribution to the systematic error; it is estimated by releasing,
in turn,  all of the background  ${\cal A}_{CP}$
parameters (limiting them within $\pm 1\sigma$ range of their measured values
for the charmless $B$ decays),
and summing in quadrature the differences
obtained from the central ${\cal A}_{CP}$ value.
A similar sum gives $^{+0.059}_{-0.057}$ as the systematic uncertainty
obtained by varying all other fixed parameters in the fit, including
$\chi_d$ and $w_l$ values, by $\pm 1\sigma$.
Finally a systematic error of $\pm 0.058$ is obtained as a result of a
null asymmetry test, when the same analysis procedure is applied
to the $B^0 \to D^{-}\rho^+$              
($D^{-} \to K^+\pi^-\pi^-$; $\rho^+ \to \pi^+\pi^0$) control sample.
To illustrate the asymmetry, we show the results separately for $\rho^0\pi^0$
candidate events tagged as $q=+1$ and $q=-1$
in Fig.~\ref{fig:rho0pi0_tagged}.

In summary, using $386 \times 10^6$ $B\overline{B}$ pairs,
we confirm evidence of $B^0 \to \rho^0 \pi^0$
decays with a branching fraction higher than most theoretical predictions
~\cite{ref:rho0pi0_predictions}. The central value remains only
slightly above the 90\% confidence-level upper limit set by
the BaBar Collaboration~\cite{ref:rho0pi0_ul}, and is in agreement
with the upper limit set by the CLEO Collaboration~\cite{ref:rhopi_bfs}.
Our measurement is consistent with, and supersedes, our
previous result~\cite{ref:rho0pi0_evidence}. 
We have also performed a first measurement of 
direct $CP$ violation in the
$B^0 \to \rho^0 \pi^0$ mode and find no statistically significant asymmetry.

The large $\rho^0\pi^0$ branching fraction suggest that one can only
impose a loose constraint on penguin uncertainty in the determination of $\phi_2$
from time-dependent  $B^0 \to \rho^\pm\pi^\mp$ measurements. It
also implies that a useful measurement of $\phi_2$ from the full $\rho\pi$
isospin analysis may be impractical even with super $B$-factory
like luminosities~\cite{ref:stark}.
Therefore, we can expect that the best measurements of
$\phi_2$ from the $\rho\pi$ system will come from the full
time-dependent amplitude analysis of $B^0 \to \pi^+\pi^-\pi^0$.

We thank the KEKB group for excellent operation of the
accelerator, the KEK cryogenics group for efficient solenoid
operations, and the KEK computer group and
the NII for valuable computing and Super-SINET network
support.  We acknowledge support from MEXT and JSPS (Japan);
ARC and DEST (Australia); NSFC and KIP of CAS (contract No.~10575109 and IHEP-U-503, China); DST (India); the BK21 program of MOEHRD, and the
CHEP SRC and BR (grant No. R01-2005-000-10089-0) programs of
KOSEF (Korea); KBN (contract No.~2P03B 01324, Poland); MIST
(Russia); MHEST (Slovenia);  SNSF (Switzerland); NSC and MOE
(Taiwan); and DOE (USA).


\begin{thebibliography}{99}
\bibitem{ref:cc}
  Throughout the paper, the inclusion of the charge conjugate mode
  decay is implied unless otherwise stated.
\bibitem{ref:average}
  The number is the average of central values reported in
  references~\cite{ref:rho0pi0_evidence,ref:rho0pi0_ul}.
\end{thebibliography}
\end{document}